\documentclass{kluwer}

\begin{document}
\begin{article}
\begin{opening}
\title{The triggered star formation in rotating disks}

\author{J. \surname{Palou\v s}\email{palous@ig.cas.cz}}
\author{S. \surname{Ehlerov\' a}\email{sona@ig.cas.cz}} 
\institute{Astronomical Institute, Academy of Sciences of the Czech Republic}
                               
\author{B. G. \surname{Elmegreen}\email{bge@watson.ibm.com}}
\institute{IBM Research Division, T. J. Watson Research Center}




\runningtitle{The triggered star formation}
\runningauthor{Palou\v s et al.}

\begin{ao}
Jan Palou\v s\\
Astronomical Institute\\
Academy of Sciences of the Czech Republic\\
Bo\v cn\' \i \  II 1401, 1401 Prague 4\\
Czech Republic
\end{ao} 


\begin{abstract} 
The gravitational instability of expanding shells triggering the formation of 
clouds and stars is analyzed. Disks with different scale-heights, 
ambient and shell velocity dispersions, mid-plane densities, rotation rates and
shear rates are explored with three dimensional numerical simulations 
in the thin
shell approximation. Three conditions for the shell collapse are specified:
the first is that it happens before a significant 
blow-out, the second requires that the shell collapses before it 
is distorted by Coriolis forces and shear, and the third requires that the
internal pressure in the accumulated gas is small and the 
fragmentation
is achieved within the expansion time. The gas-rich and slowly rotating 
galaxies are the best sites of the triggered star formation, concluding that
its importance has been much larger at the times of galaxy formation  compared
to the present epoch.    

\end{abstract}




\end{opening}
\section{Introduction}
The gravitational instability divides the ISM to fragments, 
from which the molecular clouds originate. Their subsequent subdivision
leads to the formation of stellar clusters. The question whether this chain of 
processes initiates spontaneously or whether it is triggered by an external 
push remains open. 
Probably both the spontaneous and the triggered star formation operate in 
galaxies and it is difficult to decide which is more important.
Various kinds of triggering such as a compression of pre-existing clouds,
accumulation of gas into a shell, cloud-cloud or shell-shell collisions
have been discussed by Elmegreen (1998) and Chernin et al. (1995). 
In this contribution we discuss the constrains for the gravitational 
fragmentation of expanding shells and we try to specify when and where it
operates.
\section{Model}

We use a model of a shell approximated by an infinitesimally thin
surface expanding into a stratified, non-magnetic gaseous disk  with a Gaussian
density profile of thickness $H$, mid-plane density $\rho $, velocity
dispersion $c_{ext}$, local rotation curve $V(R) \propto R^{\alpha }$, and 
local angular rotation rate $\Omega $. The shell expansion is driven 
by an energy injection rate $L$ over the time $\tau $, after which the 
energy source fades out.
At time $t$ the shell reaches the distance from the expansion center
$r$, it has the surface density $\sigma $, the local expansion speed $v$ and 
the internal velocity dispersion in the swept-up matter $c_{sh}$.

From the linear approximation of hydrodynamical and Poisson equations 
on the surface of an expanding shell Elmegreen (1994) and 
W\" unsch \& Palou\v s (2001) derived the condition for the time $t_b$
after the beginning of expansion, when the instability starts    
\begin{equation}
\omega (t_b) = -{{3v}\over{r}}+\left({{v^2} \over {r^2}}+
\left[ {{\pi G\sigma } \over {c_{sh}}} \right]^2 \right)^{1/2} = 0.
\label{eq:time1}
\end{equation}
For $t > t_b$ and $\omega (t) > 0$,
a fragmentation integral determines the time of significant collapse $t_f$:

\begin{equation} 
I_f(t_f)=\int_{t_b}^{t_f}\omega(t^\prime)dt^\prime =1. 
\label{eq:time} 
\end{equation}
Using the three-dimensional numerical simulations, 
the condition (\ref{eq:time1}) and the integral (\ref{eq:time}) are evaluated
for any part of the shell up to the time when 
$ v \le c_{ext}$ everywhere in the galaxy symmetry plane. 

\section{Results}

From many ($\sim 10 000$) models when all the above parameters were varied over an 
extended grid of values we derive the conditions of the 
gravitational instability of expanding shells: 
\begin{enumerate}
\item the disk gas surface density $\Sigma $ has to exceed some critical value
  \begin{equation}
\Sigma_{crit}  =  0.27 \left ({ E_{tot} \over 10^{51} erg} 
\right )^{-1.1} \left ({c_{ext} \over km s^{-1}}\right )^{4.1} 
10^{20} cm^{-2}, 
\label{fit}
\end{equation}
where $E_{tot} = L \times \tau$.  
The value of $\Sigma_{crit}$ depends strongly on $c_{ext}$, 
for higher values of
$c_{ext}$ fragmentation starts at higher values of $\Sigma $.
\item the ratio of the mid-plane gas density $\rho $ to the total mass density 
$\rho_{tot}$ has to be close to 1;
\item the instability parameter $Q = \kappa c_{ext} /\pi G \Sigma $ 
or analogical shear parameter $Q_A = 8^{1/2} c_{ext} A / \pi G \Sigma $ 
have to be small ($Q , Q_A \le 1 - 1.4$);
\item the ratio of the $c_{sh}^5 / (G L)$ has to be small;
\item the value $c_{sh}/c_{ext}$ has to be small. 
It may be interpreted with 
the help of the analytical solution where the critical value 
$L_{crit}$ of the 
energy injection rate can be derived (Ehlerov\' a \& Palou\v s, 2001)   
\begin{equation}
  L_{crit} = \left ( {c_{ext} \over 8.13\ kms^{-1}} \right )^4 
             \left ({c_{sh} \over kms^{-1}} \right )
             10^{51} erg Myr^{-1}.
  \label{luml2}
\end{equation} 
\noindent
If the energy injection rate $L$ is greater than $L_{crit}$ the shell fragments.
$L > L_{crit}$ transforms condition 4 into 5. 
\end{enumerate}
Note that the above conditions are interconnected: both 1 and 2 express that 
the fragmentation happens before a significant blow out to large 
$z-$distances from the
galactic plane. 4 and 5 require that $L$ has to be large and 
$c_{sh}/c_{ext}$ has to be small.
A more detailed description of the simulations and results is given by 
Ehlerov\' a \& Palou\v s (2001) and Elmegreen et al. (2001).	 

\section{Conclusions}
The gravitational instability and star formation triggered by the collapse of
an expanding shell requires the following conditions to be fulfilled: 
\begin{itemize}
\item
The gas surface density
has to surpass a critical value $\Sigma_{crit}$ given by eq. (\ref{fit}) 
and the gas mid-plane 
volume density can not be much less than the total mass mid-plane density. 
This means that if the gas represents only a small fraction of the total
mass in the disk, the other components have to be distributed in disks
of much larger thickness, producing the $K_z$ force that restricts the 
possibility of blow-out. With a  gaseous and stellar  disk of similar
thickness, the shells are gravitationally unstable if the stellar disk  
is comparable to or less massive than the gaseous one.
\item
The disk should not rotate too fast and the shear should not be too large.
High values of $A$ and  $\kappa $ increase the values of $Q_A$ and $Q$ to
the point where the shells are stable. Fast rotation and high shear distort 
the shell making large parts of it stable.
\item 
The ISM should be able to cool sufficiently fast to decrease the random 
velocities from $c_{ext}$ in the undisturbed medium to $c_{sh}$ in the 
swept-up matter, for instability $c_{sh}/c_{ext} < 0.1$ 
within the expansion time. The influence of ISM metallicity on this 
condition should be explored in the future. 
\end{itemize}
Steep dependence of $\Sigma_{crit}$ on $c_{ext}$ indicates the importance
of the self-regulating feedback: for certain $c_{ext}$ 
the star formation is triggered if $\Sigma $ surpasses certain critical value 
$\Sigma_{crit}$. Star formation is accompanied by heating of the ISM, 
increase of 
$c_{ext}$ increasing the value of $\Sigma_{crit}$ and subsequent reduction
of triggered SFR. The $\Sigma_{crit}$ for spontaneous star formation may 
be less steeply
dependent on $c_{ext}$. Consequently, at the sites where the $c_{ext}$ has been
increased, the triggered mode of star formation may have been 
suppressed and the spontaneous mode of star formation may still be 
effective at the same time.

The best place for triggered star formation are early gas rich galaxies, 
where only a small part
of the gas has been transformed to stars and where the rotation is still 
quite slow. In the present epoch the triggering is rather exceptional
restricted to high $L$ regions with enough gas and low shear. This may be 
the situation of galaxy versus galaxy collision, when the external gas
is squeezed to the central part of a galaxy. There, during the star formation
burst, the triggered star formation may operate for some time.

\acknowledgements
The authors gratefully acknowledge financial support by the Grant Agency of 
the Academy of Sciences of the Czech Republic under the grant No. 
A 3003705/1997 and the support by the grant project of the Academy of 
Sciences of the Czech Republic No. K1048102. 
BGE was supported by NSF grant AST-9870112.

\theendnotes

\end{article}

\begin{thebibliography}{}

\bibitem[1995]{Chernin}
Chernin A.D., Efremov Yu. N., Voinovich P. A. 1995, MNRAS 275, 313.

\bibitem[2001]{ehlerova}
Ehlerov\' a S., Palou\v s J. 2001, MNRAS, in press,
astro-ph/0111495

\bibitem[1994]{elmegreen1}
Elmegreen B. G. 1994, ApJ 427, 384

\bibitem[1998]{elmegreen2}
Elmegreen B. G. 1998
\newblock {\em Origins of Galaxies, Stars, Planets and Life}, ed.
C. E. Woodward, H. AQ. Thronson, M. Shull, ASP Conf. Ser. 148, p. 150.

\bibitem[2001]{elmegreen3}
Elmegreen B. G., Palou\v s J., Ehlerov\' a S. 2001, MNRAS, submitted

\bibitem[2001]{wunsch}
W\" unsch R., Palou\v s J. 2001 A\&A 374, 746
\end{thebibliography}
\end{document}